\documentclass[conference]{IEEEtran}
\IEEEoverridecommandlockouts
\usepackage{cite}
\usepackage{amsmath,amssymb,amsfonts}
\usepackage{algorithmic}
\usepackage{graphicx}
\usepackage{textcomp}
\usepackage{xcolor}
\usepackage{verbatim}
\def\BibTeX{{\rm B\kern-.05em{\sc i\kern-.025em b}\kern-.08em
    T\kern-.1667em\lower.7ex\hbox{E}\kern-.125emX}}
\begin{document}

\title{Design of optical voltage sensor based on electric field regulation and rotating isomerism electrode\\
\thanks{Identify applicable funding agency here. If none, delete this.}
}

\author{
	
\IEEEauthorblockN{ Jun Li 1, Yifan Lin 2, Nan Xie$^{\ast}$ 3}
\IEEEauthorblockA{\textit{College of Electric Engineering and Automation}, \textit{Fuzhou University}\\
  2 Xue Yuan Road, University Town, Fuzhou, Fujian 350108 \\
$^{\ast}$t13056@fzu.edu.cn}
}
\maketitle

\begin{abstract}

Temperature drift, stress birefringence and low frequency vibration lead to the randomness and fluctuation of the output of optical voltage sensor(OVS). In order to solve the problem, this study adopts the lock-in amplifier technology with the aid of a high-speed rotating electrode to realize electric field modulation. This technology could shift the measured signal frequency band from near 50 Hz moved to several kilometer Hz, so as to make the output signal avoid the interference from low-frequency temperature drift, stress birefringence and vibration, leading to higher stability and reliability.
The electro-optic coupling wave theory and static electric field finite element method are utilized to investigate the shape of modulation wave.  The simulation results proves that lock-in technology is able to prevent the measured voltage signal from the large step signal interference and restore the perfect original signal. While the sample rate is decreased to the value of the modulation frequency.

\end{abstract}


\section{Introduction}

Optical voltage sensor (OVS) based on The Pockels effect has the advantages of small volume, good insulation performance, no ferromagnetic saturation, large dynamic measurement range and high bandwidth, and is the development direction of primary measurement equipment for smart grid [1,2].  However, its long-term stability is poor and it has not been widely used in power system.  It is generally believed that temperature, stress, linear birefringence and vibration, aging are the main factors affecting the long-term operation stability of OVS, and these factors are coupled with each other [3-8].

In order to solve the above problems, the research groups all over the world has put forward a variety of optimization methods, which are discussed as following categories:

\paragraph{Stress birefringence suppression}
Dual optical path[3] and dual crystal method[4] are always utilized to suppress the linear birefringence. Since the linear birefringence has the charactors of random and fluctuaion. The method could not fully eliminate the interference.

\paragraph{Temperature drift compensation}
 Bohnert[5] et al proposed utilizing the temperature drift of the  dielectric coefficient to compensate the one of electro-optic coefficient by using quartz voltage divider. Also, software temperature compensation method is often utilized with the help of measuring the crystal's temperature in real time[6]. 

\paragraph{phase modulation with closed loop control}

For all fiber optical current sensor(AFOCS)[7,8],
by adding the phase delay modulation with square wave shape
the static operating point would be always shifted to the linear region. However, the closed-loop control is needed because the random drift is always exist. However, when temperature drift is alter dramatically, the PID parameters would be unreliable, and the system would be unstable.

This paper proposes a method based on rotating heterogeneous electrode which modulated the electric field's direction and magnitude. With the aid of DSP, the digital lock-in amplifier could be implemented to restore the original signal which would be disturbed by the temperature drift, vibration and linear birefringence. A simulation conducted by Simulink has proved the feasibility of the rotated-electrode OVS. The paper is organized as follows: Section II describe the structure of the sensor, and briefly discuss the function of several elements. Section II discuss the modulation wave shape results from the rotation electrode. Section III calculate the reference signal. Section IV shows the principle of Lock-in amplifier in the case that the modulation and reference signal are neither sinusoidal wave or square one. Section V show the simulation results with the aid of Simulink. Section VI give the conclusion.

\section{constituents of sensor}

As shown in Fig 1, the whole system consists of rotating electrode, high voltage electrode, BGO crystal, polarized optical elements, semiconductor light source, Photo diode, motor, digital signal processor (DSP) and so on. The first five components constitute an conventional optical voltage sensor based on Pockels effect. Since the quarter wave plate (QWP) shifts the working zero point to the linear region, the change of output light intensity is proportional to the applied voltage and the electric field inside the crystal. Electrical field modulation and lock-in amplifier system are composed of rotating ground electrode, motor, LED, PD and DSP. Due to the rotation of ground electrode, the direction and magnitude of field inside BGO crystal are altered in period, leading to the modulation of output light power. By utilizing digital lock-in amplifier technology with DSP,
the effective signal is displaced from the noisy low-frequency band(~50Hz) to the middle-low frequency (~2kHz) band, avoiding the disturb from coupling of linear birefringence, temperature and vibration.

\begin{figure}[htbp]
	\centerline{\includegraphics[width=0.35\textwidth]{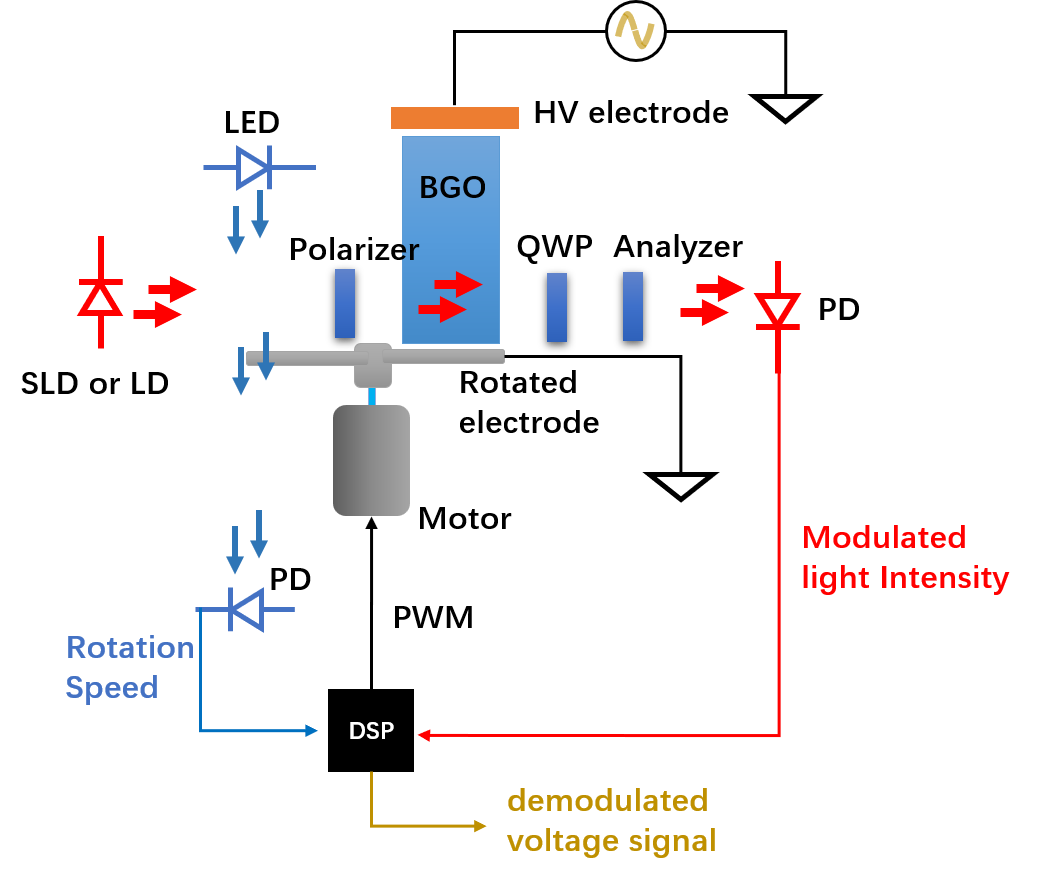}}
	\caption{The schematic of experimental set-up.}
	\label{fig}
\end{figure}

The rotation speed of the hollow-cup DC-motor and three-phase brush-less motor could reach 30,000-150,000 round per minute(RPM), that is, the rotation frequency of 500-2500Hz. If the rotated electrode adopts multi-blade structure, the modulation frequency would increase up to 10kHz, whose bandwidth is sufficient to detect the switch noise raised from the accessed distributed power devices. 

The light switch composed of LED and PD monitors the speed of motor. It grantee the constant speed of the motor and thus the constant modulation frequency. On the other hand, it provide a reference signal for the demodulation of phase-locked amplifier.

The key points to be investigated in the system are: 

\textit{1. The waveform of modulated light intensity.}  Due to the shape of electrode's blade, the shape of wave would deviates from the perfect sinusoidal signal and contains high harmonic component.

 \textit{2. The waveform of the reference signal. }Because of the LED light's divergence, the light spot on electrode's blade usually has a certain area, which may larger than the blade itself. And the final received  reference signal is neither an ideal sinusoidal signal nor a square wave signal, need to be studied in detail. 
 
 \textit{3. Analysis of the lock-in amplifier's principle.} Since neither the modulation signal nor the reference signal is an ideal sinusoidal signal, corresponding analysis is required. 

The three aspects will be discussed in the following sectors.


\section{performance of rotated electrode}

The structure of OVS sensor-head with rotating electrode is shown in the Figure 2. The optical path for measurement of electric field is perpendicular to the one for reference signal (rotation frequency measurement).  The strip region in the BGO crystal is where light path is located and the finite element grids is denser than adjacent area for investigating precise change of the electric field. The rotating electrode is a mono 30-degree fan-shaped blade, the rotation axis is shown as a black arrow in Fig 2.
\begin{figure}[htbp]
	\centerline{\includegraphics[width=0.25\textwidth]{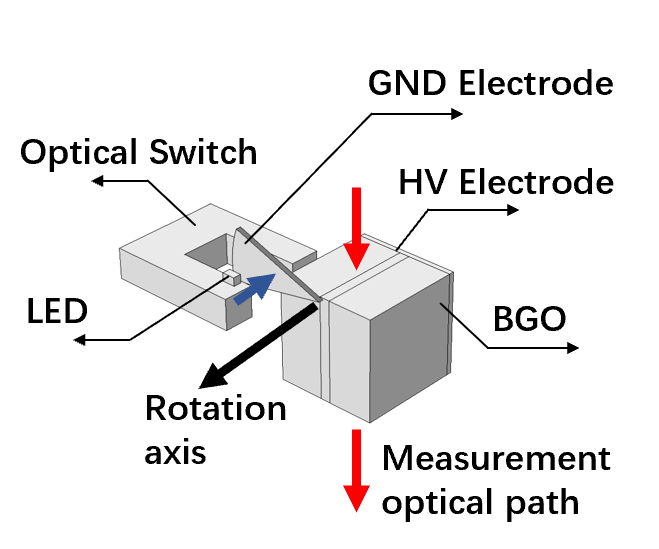}}
	\caption{The structure of rotated electrode sensor head}
	\label{fig}
\end{figure}

 As shown in Fig 3, When the blade runs directly covering the light-pass region, it is equivalent to the conventional OVS structure. In this position, the potential gradient in the crystal changes  most, and the electric field and optical signal alters the most. When the blade rotates to a position far away from the crystal, electric potential changes in the crystal are small, the direction of the electric field is inclined to 45 degrees, and the electric field is the smallest and the optical signal is the lowest. 

\begin{figure}[htbp]
	\centerline{\includegraphics[width=0.28\textwidth]{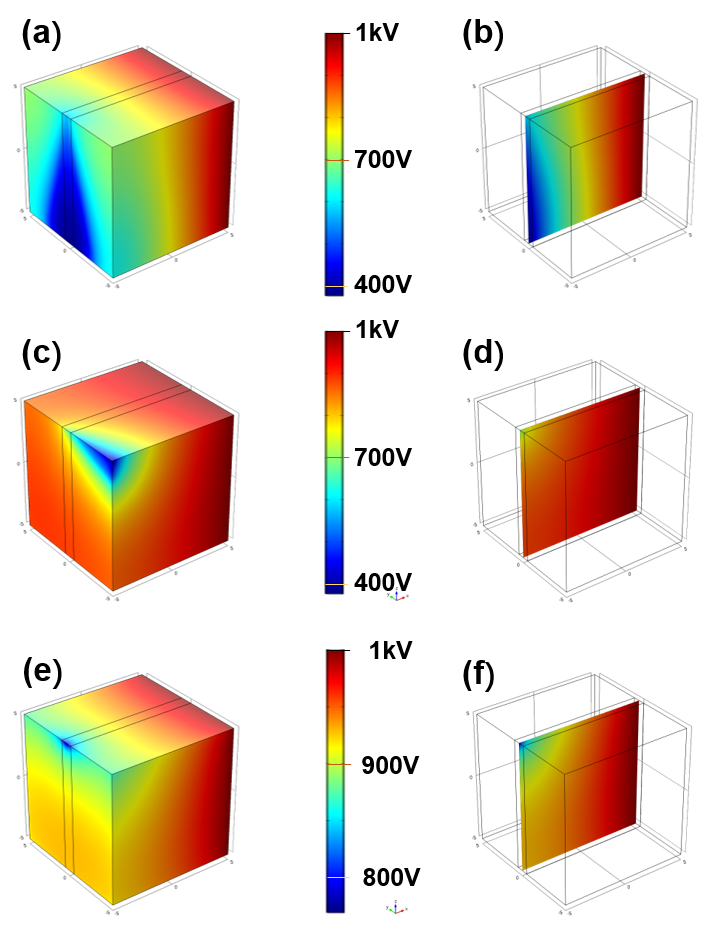}}
	\caption{The voltage potential simulation of BGO crystal for different position of rotated electrode.}
	\label{fig}
\end{figure}

\begin{figure}[htbp]
	\centerline{\includegraphics[width=0.3\textwidth]{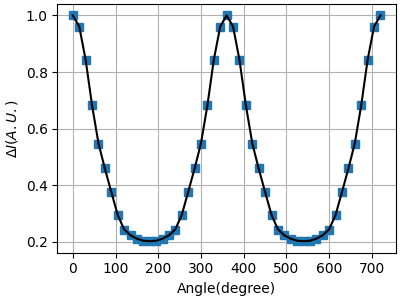}}
	\caption{The waveform of modulated signal.}
	\label{fig}
\end{figure}
The electro-optic coupling wave theory was utilized to simulate the output optical signal at different positions of electrodes. And their relationship is depicts in Fig 4, which shows two complete cycles, namely 720 degrees. Fourier series fitting was carried out, and the fit function is shown in equation (1).

\begin{equation}
	f(\alpha) =\sum_{i=1}^7 A_i \cos{(i \alpha + \Phi)}
\end{equation} 
 The fitting coefficients are shown in the Table I. It could be found that the modulated light intensity is not a perfect sinusoidal signal, and the amplitude of the second harmonic wave is of the same magnitude as that of the fundamental wave.
\begin{table}[htbp]
	\renewcommand{\arraystretch}{1.3}
	\caption{Fitting parameter of modulated wave}
	\label{table_example}
	\centering
	\begin{tabular}{c c c c c}
		\hline
		$\bf{ A_1}$ & $\bf{ A_2}$ & $\bf{A_3}$ &$\bf{A_4}$ & $\bf{ A_5}$\\
	
		$\bf{ A_6}$ & $\bf{A_7}$ & $\bf{\Phi}$ & \bf{c}& \\
		 \hline\hline
		  0.366 & 0.118&  0.032 & 0.018&
		 5.4e-3\\

		 -6.2e-3& -4.3e-3& -2.4e-5&
		0.471&\\
		\hline
	\end{tabular}
\end{table}

\section{generation of reference signal}

Reference signal can be generated by a variety of ways, its principle  is similar to that of encoder of three-phase brushless DC motor, including Hall measurement, inverse electrodynamic potential measurement, photoelectric type and so on. In this paper, the photoelectric signal generation method is applied. Thanks to the optical switch containing LED and Photo-diode, The rotation speed measurement could be conducted easily and the reference signal is provided.

According to the instruction of the optical switch, the relation between the emitted light intensity and the emission angle is shown in the Figure A1, which can be fitted by formula 2,
\begin{align}
	I(\beta) =& A\cos{(k\beta)} + c
\end{align}
where $\beta$ is the emission angle of LED, A,k and c is the fitting parameter. 
the fitting results is A = 4.113, k = 0.0789*180/$\pi$, c=4.227.
Its shape is a better cosine function waveform.


Suppose that the distance between LED and the center of rotating electrode is d, and the distance between the rotation center and the center of LED spot is R0, and the distance between the calcuated point and the spot center is r, which is related to the emission angle $\beta$. we have 
\begin{align}
	\beta =& \arctan{(r/d)}
\end{align}

For large spot, that is, the spot area cannot be completely blocked by the blade,as shonw in Fig A2. the expression of the light intensity pass-through is shown in equation (A1). 
and the bow function is 

\begin{align}
	I_{bow}^{\mp}(\alpha) =& \int_{\pm r_0 \cos{\alpha}}^{ r_0 } \int_{- \sqrt{r_0^2-x^2}}^{\sqrt{r_0^2-x^2}} I(\beta) dxdy\\
	r =& \sqrt{x^2+y^2}\\
		\alpha =& \arccos{(R_0 \sin{\theta} / r_0)}
\end{align}
where

In the actual situation, the receiving surface of PD is very small, only a few mm, which is equivalent to the small area of the effective spot. At this point, the spot will be completely blocked by the sector area, and its expression is

\begin{equation}
	{I_{out}} = \begin{cases}
		1, &{\text{if}}\ -\pi<\theta<-\theta_{\text{max}}  \\
		I_{\text{bow}}^{+}(\theta),&{\text{if}}\ -\theta_{\text{max}}<\theta<0  \\ 
		{I_{\text{bow}}^{-}(\theta),}&{\text{if}}\ 0<\theta<\theta_{\text{max}}\\
		0, &{\text{if}}\ \theta_{\text{max}}<\theta<\theta_{\text{GND}}/2
	\end{cases}
\end{equation}
where
\begin{align}
	\theta_{\text{max}} =& \arcsin{(r_0/R_0)}
\end{align}

The calculated results is shown if Fig. 5, and the paramiter is   $r_0$ = 0.5mm, $d$ = 2mm, the angle of the electrode with sector shape is $\theta_{\text{GND}}$ = 30 degree, and $R_0$ = 6mm.

\begin{figure}[htbp]
	\centerline{\includegraphics[width=0.27\textwidth]{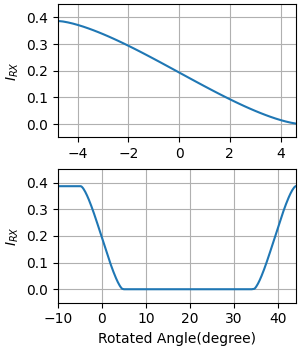}}
	\caption{ Calculated reference signal generated from a optical switch.}
	\label{fig}
\end{figure}

It can be seen from the above results that if PD with large receiving area is adopted and multi-blade design of ground electrode is adopted, the reference signal is likely to be an ideal sinusoidal signal. However, the cost of the large area PD or four quadrant detector is higher. In this paper, the actual output is a  trapezoidal wave, but the duty cycle is too large to be directly used as reference signals. However the digital square and sinusoidal signals used for demodulation could be generated from the real-time value of trapezoidal wave's period.

\section{principle of Lock-in Amplifier}






The digital phase-locked amplification technique has been described in many literatures[9-11]. 
the product of  modulated signal and the reference sinusoidal signal is integrated within a modulation period to obtain the original signal, and the rate of amplitude and phase delay between the modulating signal plus the original signal and the reference are calculated.

The above technology generally consider both the modulating signal and the reference one to be sine wave. However, it is not true in this paper, thus this paper mainly discuss scenario that both the modulation and reference signals are neither perfect sinusoidal wave nor square one, and contains considerable higher harmonics.


Lock-in amplifier technology is able to measure any waveform, including the signal of AC bus system or DC bus system. Suppose the voltage signal to be measured as S(T). and the modulation signal is M(t). and the modulated signal would be written as
\begin{align}
	S_{m}(t) = M(t) S(t) 
\end{align}
where M is a period function whose wave is shown in Fig 4, and could   be expand to Fourier series as follows,
\begin{align}
	M(t) = m_{0}+ \sum_{j=1}^{l} m_x^j\cos (2 \pi jf_m t ) 
	+ m_y^j\sin (2 \pi jf_m t ) 
\end{align}

Suppose the reference signal is not perfect square or sinusoidal, could also  be expand to Fourier series,
\begin{equation}
	R(t) = m_{0} + \sum_{k=1}^{l} r_x^k\cos (2 \pi kf_m t )
	+ r_y^k\sin (2 \pi kf_m t )
\end{equation}
In another respective, R could be considered as the combination of odd and even components, and
\begin{align}
	R_{odd} =&  \sum_{j=0}^{l}  m_x^j\sin 2 \pi( j-1)f_m  t \\
	R_{even} =&  \sum_{j=0}^{l}  m_x^j\cos 2 \pi( j-1)f_m  t 
\end{align}
In DSP or MCU, the even component of R is multiplied with modulated signal $S_m(t)$,
\begin{equation}
	\begin{aligned}
		S_m(t)&R_{even}(t) \\ 
		= & S(t)\sum_{j=0}^{l}\sum_{k=0}^{l}
		m_x^jr_x^k\cos (2 \pi jf_m t)\cos (2 \pi kf_m t )\\
		&+m_y^jr_x^k\sin(2 \pi jf_m t)\cos (2 \pi kf_m t )
	\end{aligned}
\end{equation}
Integrate both side of equation 16 within the time of $T_m = 1/f_m$ ,and note that
\begin{align}
	\int_{0} ^{T_m}\cos(2\pi jf_m t)\cos(2\pi kf_k t) dt &=\delta_{mk}	\\
	\int_{0} ^{T_m}\sin(2\pi jf_m t)\sin(2\pi kf_k t) dt &=\delta_{mk}	\\
	\int_{0} ^{T_m}\sin(2\pi jf_m t)\cos(2\pi kf_k t) dt &=0
\end{align}

we get 
\begin{equation}
	\int_{0} ^{T_m} S_m(t)R_{even}(t) = S(t)\sum_j^l m_x^j r_x^j 
\end{equation}
and 
\begin{equation}
	S(t) = \frac{\int_{0} ^{T_m} S_m(t)R_{even}(t)}{\sum_j^l m_x^j r_x^j } 
\end{equation}
also we get
\begin{equation}
	S(t) = \frac{\int_{0} ^{T_m} S_m(t)R_{odd}(t)} {\sum_j^l m_y^j r_y^j }
\end{equation}

Especially, for modulated wave S(t)M(t), the amplitude and phase delay for every harmonic wave could be written as

\begin{align}
	\begin{aligned}
		X_{out}^i =&S(t)m_x^i
		\\=& \frac{m_x^i}{\sum_j^l m_x^j r_x^j}\int_{0} ^{T_m} S_m(t)R_{even}(t) dt
	\end{aligned}
\end{align}
and
\begin{align}
	\begin{aligned}
		Y_{out}^i =&S(t)m_y^i
		\\=& \frac{m_y^i}{\sum_j^l m_y^j r_y^j}\int_{0} ^{T_m} S_m(t)R_{odd}(t) dt	
	\end{aligned}
\end{align}
So the magnitude and phase of the \textit{i}th harmonic wave is
\begin{align}
	M_{out}^i=&\sqrt{(X_{out}^i)^2+(Y_{out}^i)^2}\\
	A_{out}^i=&\arctan{(X_{out}^i/Y_{out}^i)}
\end{align}

\section{simulation of system}

The Simulink model  consist of four parts in total, as shown in Figure 6. The left-upper part is the reference signal(R(t)) module, which could be selected from 2.5kHz square wave or 2.5kHz sinusoidal wave. The left-center is modulation module. The signal wave S(t) is set to be a 50Hz sinusoidal signal with an amplitude of 1. The modulation waveform M(t) is shown in Figure 4 with a frequency of 2.5kHz as well. The two signal are multiplied to produce the modulated signal S(t)M(t). The left-lower module is the noise (N(t)) module, which could be selected between the random square wave (white noise module) and the 10Hz sinusoidal wave with amplitude of 10. The former represent random step signal in DC measurement due to the polarization or discharge, and the latter simulates  random and fluctuating disturb signal results from the linear birefringence in optical fiber. 

\begin{figure}[htbp]
	\centerline{\includegraphics[width=0.4\textwidth]{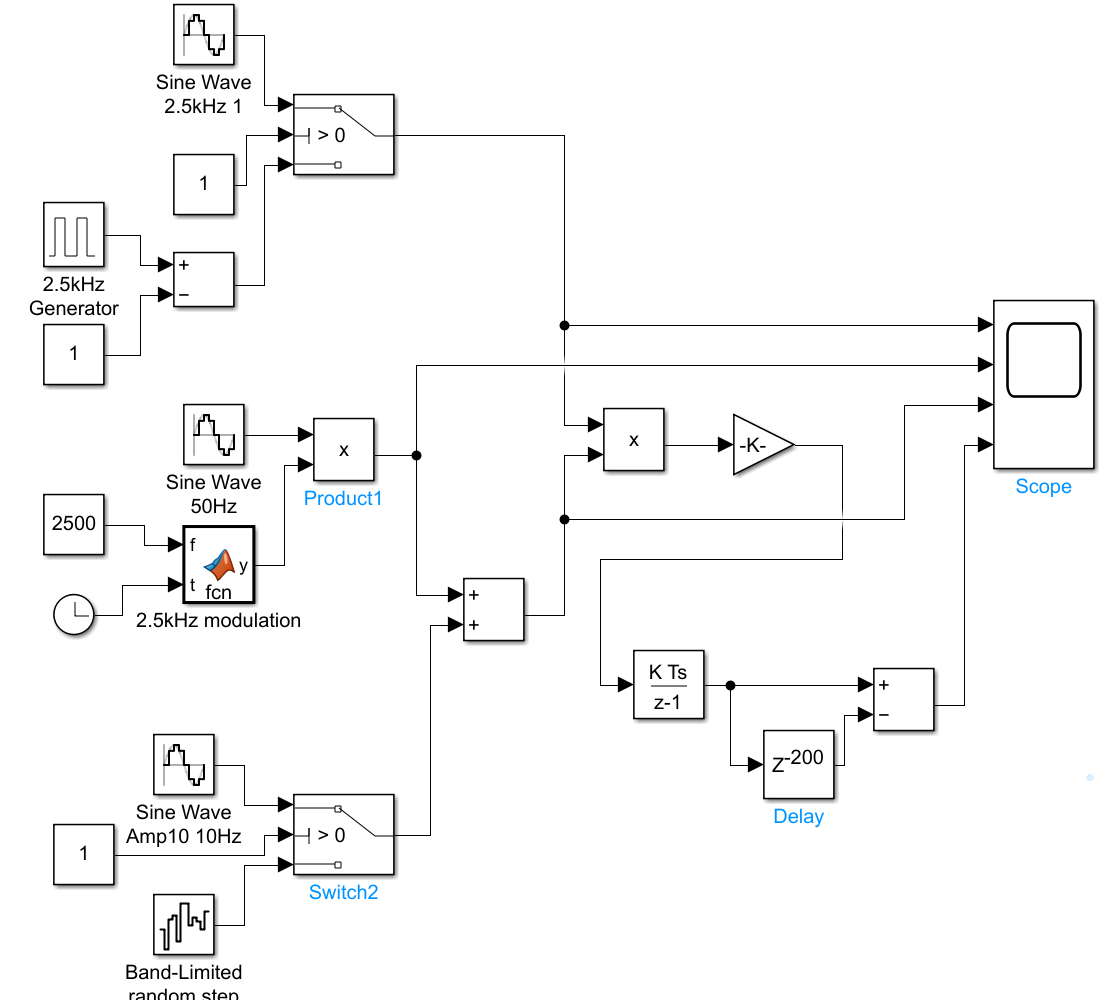}}
	\caption{Simulation model of Lock-in for rotation electrode sensor}
	\label{fig}
\end{figure} 

the final signal is the modulated signal S(t)M(t) added the noise N(t). To recover the original signal S(t), S(t)M(t) is multiplied with R(t), and then integrate within a time of $T_m$. The definite integral module consists of two integral module with a phase delay of $T_m$, and their subtraction is the needed output.  

The simulation parameter is set as follows, the time step is 2e-6 s, and the time delay of the two integration model is set 200 time steps, namely the period of the reference signal. And Running time is set to 0.03 second. The reference signal has a phase delay of $\pi/6$ compared to the modulation signal.


\begin{figure}[htbp]
	\centerline{\includegraphics[width=0.35\textwidth]{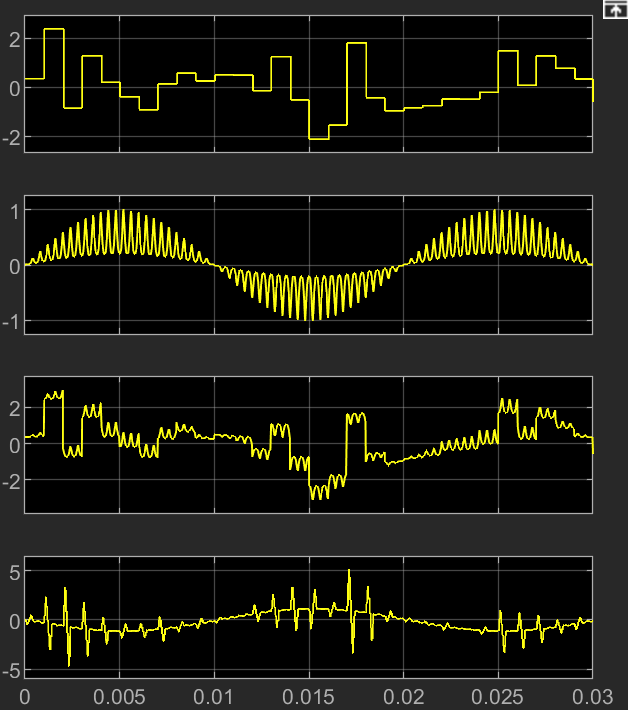}}
	\caption{Simulation results.}
	\label{fig}
\end{figure} 
Simulink results are shown in Figure 7, from top to bottom, the first signal represents the step noise caused by discharge and polarization, which is independent of the measured voltage. The second one is a modulated 50Hz signal with an amplitude of 1. The third is the modulated signal after adding interference, and the characteristics of original signal would hardly be observed. The fourth is restored signal after integration, and the spike  is generated at the step point of the noise.

To further suppress the spike of restored signal, it is down-sampled at a frequency of 2.5kHz, that is, the reference signal frequency. And Only the points with modulation phase of 0 were retain to obtain the perfect original signal, as shown in the figure 8. The red curve shows the perfect restore signal with 180 degree delay, which has been artificially shifted up by 3 for ease of observation. The disadvantage of down-sampling is that the actual sampling rate is equivalent to the frequency of modulated signal. According to Nyquist's Theorem, the effective bandwidth is only 1.25kHz.

\begin{figure}[htbp]
	\centerline{\includegraphics[width=0.35\textwidth]{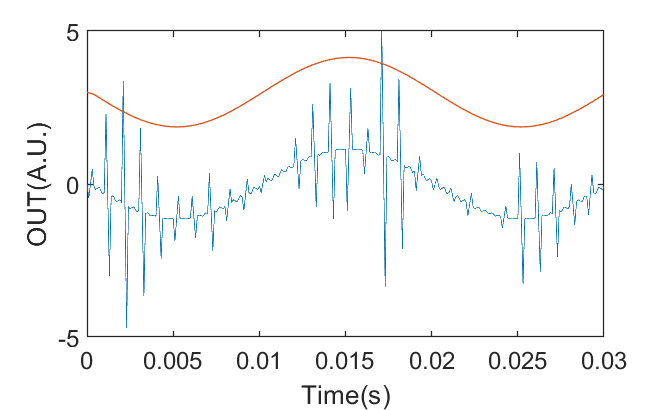}}
	\caption{Restored signal(blue) and its down-sampling signal(red), the latter is shift up of 3.}
	\label{fig}
\end{figure}



\section{conclusion}


In this paper, a rotated-electrode optical voltage sensor is designed to suppress low-frequency birefringence and eliminate the interference of temperature and low-frequency vibration on the measurement results. A DC motor is used to drive the ground electrode to rotate in its plane, so as to modulate the direction of the electric field and get the modulation optical signal related to the measured voltage. The signal is demodulated using lock-in technology, and the scheme is verified by Simulink. The restoration of the original waveform needs to reduce the actual sampling number of the signal,namely the modulation frequency of the signal in this paper. The methods to further improve the sampling number and resolution are increasing the number of ground electrode blades, increasing the rotational speed and using modulation frequency dither technology[11].

\section{ACKNOWLEDGMENT}
The author would like to thank Prof. Xinghua Lu in Institute of Physics, Chinese Academic of Science for the valuable and inspired discussion. The work is supported by National Nature Science Foundation of China(No. 51807030).

\cleardoublepage
{\appendices
	\renewcommand\thefigure{\Alph{section}\arabic{figure}} 
	\renewcommand\thetable{\Alph{section}\arabic{table}}
	\renewcommand\theequation{\Alph{section}\arabic{equation}}

\section{supplement for generation of Reference}
\setcounter{figure}{0} 
\setcounter{equation}{0} 

Fig A1 shows the LED's emission intensity versus emission angle.
\begin{figure}[htbp]
	\centerline{\includegraphics[width=0.23\textwidth]{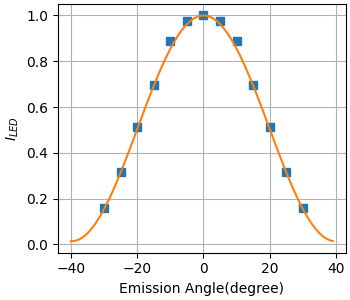}}
	\caption{The architecture of replaced message}
	\label{fig}
\end{figure}

Figure A2 shows the process how the LED spot is sheltered by the blade.

\begin{figure}[htbp]
	\centerline{\includegraphics[width=0.4\textwidth]{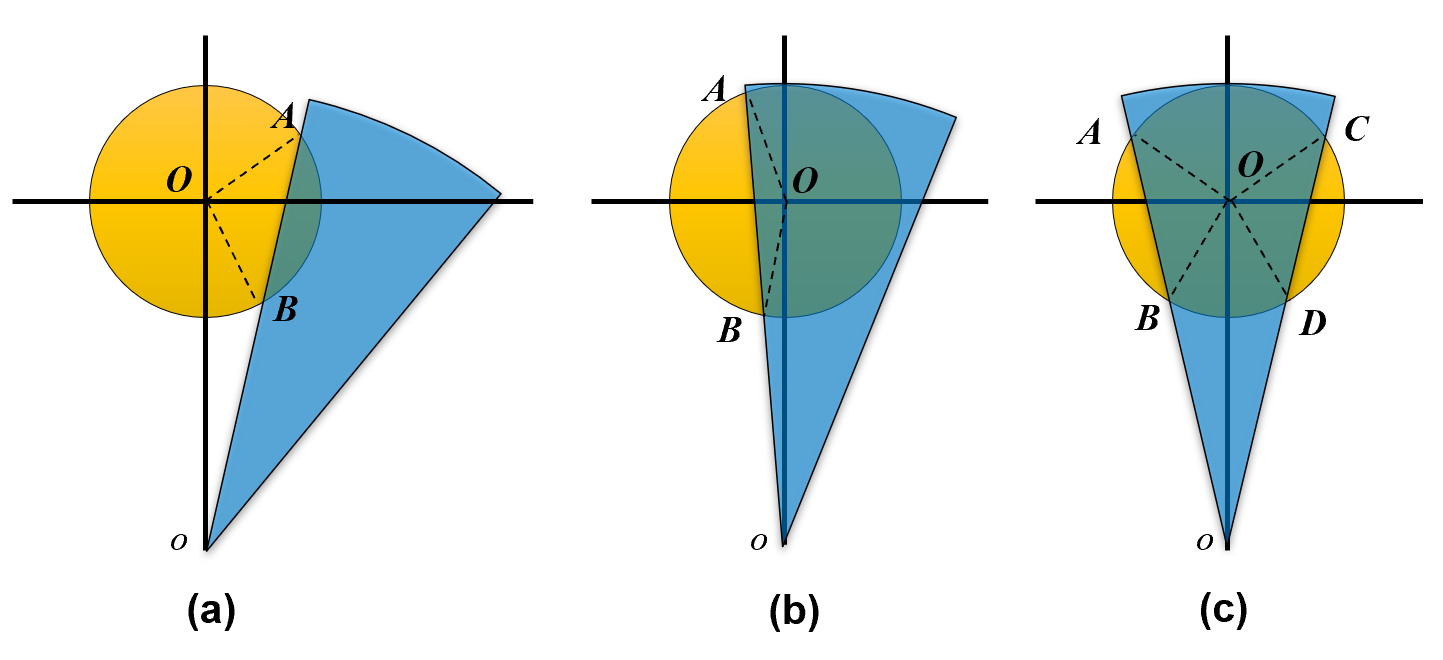}}
	\caption{The architecture of replaced message}
	\label{fig}
\end{figure}

Equation describe the process of sheltering as shown in Figure A2.
\begin{equation}
	{I_{out}} = \begin{cases}
		1, &{\text{if}}\ \theta<-\theta_{\text{max}}  \\
		I_{\text{bow}}^{+}(\theta),&{\text{if}}\ -\theta_{\text{max}}<\theta<0  \\ 
		{I_{\text{bow}}^{-}(\theta),}&{\text{if}}\ 0<\theta<\theta_{\text{GND}}- \theta_{\text{max}}\\
		{I_{\text{bow}}^{-}(\theta)+I_{\text{bow}}^{-}(\theta'),}&{\text{if}}\ 0<\theta<\theta_{\text{GND}}-\theta_{\text{max}}
	\end{cases}
\end{equation}

Figure A3 shows the transmitted light intensity with rotating electrode verus the rotation angle.
\begin{figure}[htpb]
	\centerline{\includegraphics[width=0.25
		\textwidth]{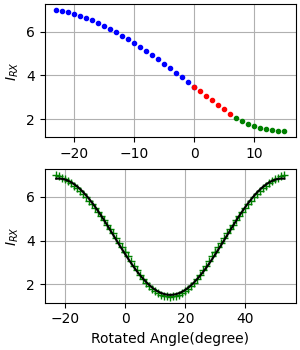}}
	\caption{The architecture of replaced message}
	\label{fig}
\end{figure}

Figure A3 could be fitted as Equation A2, And the parameter is B = 2.653, u = -4.8316, $\Phi$ = 4.4065, $c_2$ = 4.1773.
\begin{align}
	I_{out} =& B\cos{ (u \theta + \Phi)} + c_2
\end{align}

\end{document}